\documentclass[conference]{IEEEtran}

\usepackage[T1]{fontenc}
\usepackage{blindtext}
\usepackage[utf8]{inputenc}
\usepackage[noadjust]{cite}
\usepackage[font=footnotesize]{caption}
\usepackage{dblfloatfix}
\usepackage{flafter}
\usepackage{graphicx}
\usepackage{psfrag}
\usepackage{subfig}
\usepackage{amsmath,amssymb,amsthm,mathrsfs,amsfonts,dsfont}
\usepackage{color}
\usepackage[algoruled]{algorithm2e}
\usepackage{fixltx2e}
\usepackage{multirow}
\usepackage{multicol}
\usepackage{nohyperref}  

\usepackage[normalem]{ulem}
\usepackage{acronym}
\usepackage[table,xcdraw]{xcolor}
\usepackage{accents}
\usepackage{mwe}
\usepackage{hhline}
\usepackage{trfsigns}
\usepackage{siunitx}
\usepackage{lscape}
\usepackage{wasysym}
\usepackage{bm}
\usepackage{xfrac} 
\usepackage{url}

\makeatletter
\newcommand*\bigcdot{\mathpalette\bigcdot@{.5}}
\newcommand*\bigcdot@[2]{\mathbin{\vcenter{\hbox{\scalebox{#2}{$\m@th#1\bullet$}}}}}
\makeatother

\usepackage{tikz}

\definecolor{nokiablue}{rgb}{0, 0.3529, 1}
\definecolor{nokiapurple}{rgb}{0.5020, 0.4314, 0.9490}

\newacro{3GPP}{3rd Generation Partnership Project}
\newacro{5G}{fifth generation}
\newacro{5G NR}{Fifth Generation New Radio}
\newacro{6G}{sixth generation}
\newacro{A/D}{analog-to-digital}
\newacro{ABE}{analog back-end}
\newacro{ADC}{analog-to-digital converter}
\newacro{AFE}{analog front-end}
\newacro{AGV}{automatic guided vehicle}
\newacro{AWGN}{additive white Gaussian noise}
\newacro{B5G}{beyond \ac{5G}}
\newacro{BB}{baseband}
\newacro{BER}{bit error rate}
\newacro{BLER}{block error rate}
\newacro{BPSK}{binary phase-shift keying}
\newacro{BP}{band-pass}
\newacro{BS}{base station}
\newacro{CDM}{code-division multiplexing}
\newacro{CFO}{carrier frequency offset}
\newacro{CFR}{channel frequency response}
\newacro{CIR}{channel impulse response}
\newacro{CoMP}{coordinated multipoint}
\newacro{CP}{cyclic prefix}
\newacro{CPE}{common phase error}
\newacro{CPO}{carrier phase offset}
\newacro{CRLB}{Cram\'er?Rao lower bound}
\newacro{CS}{chirp sequence}
\newacro{CSI}{channel state information}
\newacro{CW}{continuous wave}
\newacro{CZT}{chirp Z-transform}
\newacro{D/A}{digital-to-analog}
\newacro{DS2D}{direct satellite-to-device}
\newacro{DAC}{digital-to-analog converter}
\newacro{DDS}{direct digital synthesis}
\newacro{DFRC}{dual-function radar-cunication or dual-functional radar-cunication}
\newacro{DFnT}{discrete Fresnel transform}
\newacro{DFT}{discrete Fourier transform}
\newacro{DMRS}{demodulation reference signal}
\newacro{DoA}{direction-of-arrival}
\newacro{DoD}{direction-of-departure}
\newacro{DPD}{digital pre-distortion}
\newacro{ETSI}{European Telecommunications Standards Institute}
\newacro{EVM}{error vector magnitude}
\newacro{FDE}{frequency-domain equalization}
\newacro{FDM}{frequency-division multiplexing}
\newacro{FO}{frequency offset}
\newacro{FR1}{Frequency Range 1}
\newacro{GEO}{geostationary Earth orbit}
\newacro{gNB}{gNodeB}
\newacro{HP}{high-pass}
\newacro{HPBW}{half-power beamwidth}
\newacro{IBFD}{in-band full duplex}
\newacro{ICI}{intercarrier interference}
\newacro{IDFT}{inverse discrete Fourier transform}
\newacro{IDFnT}{inverse discrete Fresnel transform}
\newacro{IF}{intermediate frequency}
\newacro{IHE}{Institute of Radio Frequency Engineering and Electronics}
\newacro{I/Q}{in-phase/quadrature}
\newacro{ISAC}{integrated sensing and communication}
\newacro{ISI}{intersymbol interference}
\newacro{ISLR}{integrated-sidelobe level ratio}
\newacro{IoT}{Internet of Things}
\newacro{JCAS}{joint communication and sensing}
\newacro{KIT}{Karlsruhe Institute of Technology}
\newacro{LDPC}{low-density parity-check}
\newacro{LEO}{low Earth orbit}
\newacro{LFSR}{linear-feedback shift register}
\newacro{LNA}{low-noise amplifier}
\newacro{LO}{local oscillator}
\newacro{LoS}{line-of-sight}
\newacro{LP}{low-pass}
\newacro{LS}{least squares}
\newacro{MEO}{medium Earth orbit}
\newacro{mmWave}{milimeter wave}
\newacro{MIMO}{multiple-input multiple-output}
\newacro{MLE}{maximum likelihood estimator}
\newacro{MLS}{maximum-length sequence}
\newacro{MRC}{maximal-ratio combining}
\newacro{MUSIC}{multiple signal classification}
\newacro{NB}{narrowband}
\newacro{NLoS}{non-line-of-sight}
\newacro{NR}{new radio}
\newacro{NTN}{non-terrestrial networks}
\newacro{OCDM}{orthogonal chirp-division multiplexing}
\newacro{OFDM}{orthogonal frequency-division multiplexing}
\newacro{OOB}{out-of-band}
\newacro{OTA}{over-the-air}
\newacro{P/S}{parallel-to-serial}
\newacro{PA}{power amplifier}
\newacro{PACF}{periodic autocorrelation function}
\newacro{PCCF}{periodic cross-correlation function}
\newacro{PLC}{powerline cunication}
\newacro{PLL}{phase-locked loop}
\newacro{PMCW}{phase-modulated continuous wave}
\newacro{PMN}{perceptive mobile network}
\newacro{PN}{oscillator phase noise}
\newacro{PoC}{proof-of-concept}
\newacro{PPLR}{peak power loss ratio}
\newacro{PRBS}{pseudorandom binary sequence}
\newacro{PRS}{positioning reference signal}
\newacro{PSLR}{peak-to-sidelobe level ratio}
\newacro{QPSK}{quadrature phase-shift keying}
\newacro{RadCom}{radar-cunication}
\newacro{RCS}{radar cross section}
\newacro{RF}{radio-frequency}
\newacro{RIS}{reflective intelligent surface}
\newacro{RMSE}{root mean squared error}
\newacro{SAC}{synthetic aperture communication}
\newacro{SAR}{synthetic aperture radar}
\newacro{SATCOM}{satellite communication}
\newacro{SC}[S\&C]{Schmidl \& Cox}
\newacro{SFO}{sampling frequency offset}
\newacro{SIC}{self-interference cancellation}
\newacro{SINR}{signal-to-interference-plus-noise ratio}
\newacro{SIR}{signal-to-interference ratio}
\newacro{SISO}{single-input single-output}
\newacro{SNR}{signal-to-noise ratio}
\newacro{SoC}{system-on-a-chip}
\newacro{SSB}{synchronization signal block}
\newacro{STO}{symbol time offset}
\newacro{S/P}{serial-to-parallel}
\newacro{TDE}{time-domain equalization}
\newacro{TDM}{time-division multiplexing}
\newacro{TDR}{time-domain reflectometry}
\newacro{TITO}{tilt inference of time offset}
\newacro{TO}{time offset}
\newacro{UE}{user equipment}
\newacro{UL}{uplink}
\newacro{ULA}{uniform linear array}
\newacro{URA}{uniform rectangular array}
\newacro{UWAC}{underwater acoustic communication}
\newacro{V2V}{vehicle-to-vehicle}
\newacro{VENUS}{virtually enhanced satellite uplink via receiver aperture synthesis}
\newacro{ZF}{zero forcing}
\newacro{ZP}{zero padding}

\makeatletter
\renewcommand*\env@cases[1][1.2]{%
	\let\@ifnextchar\new@ifnextchar
	\left\lbrace
	\def\arraystretch{#1}%
	\array{@{}l@{\quad}l@{}}%
}
\makeatother

\usepackage{fancyhdr}
\fancypagestyle{empty}{fancy}

\begin{document}

\title{Synthetic Aperture Communication: Principles and Application to  Massive IoT Satellite Uplink}

\author{\IEEEauthorblockN{Lucas Giroto, Marcus Henninger, and Silvio Mandelli}\\
		\IEEEauthorblockA{Nokia Bell Labs Stuttgart, 70469 Stuttgart, Germany \\
			E-mail: \{firstname.lastname\}@nokia-bell-labs.com
		}
	}

\maketitle

\begin{abstract}
    While synthetic aperture radar is widely adopted to provide high-resolution imaging at long distances using small arrays, the concept of coherent synthetic aperture communication (SAC) has not yet been explored. This article introduces the principles of SAC for direct satellite-to-device uplink, showcasing precise direction-of-arrival estimation for user equipment (UE) devices, facilitating spatial signal separation, localization, and easing link budget constraints. Simulations for a low Earth orbit satellite at \mbox{\SI{600}{\kilo\meter}} orbit and two UE devices performing orthogonal frequency-division multiplexing-based transmission with polar coding at \mbox{\SI{3.5}{\giga\hertz}} demonstrate block error rates below 0.1 with transmission powers as low as $\SI{-10}{dBm}$, even under strong interference when UE devices are resolved but fall on each other's strongest angular sidelobe. These results validate the ability of the proposed scheme to address mutual interference and stringent power limitations, paving the way for massive Internet of Things connectivity in non-terrestrial networks.
\end{abstract}

\acresetall



\IEEEpeerreviewmaketitle


\section{Introduction}\label{sec:introduction}
Direct satellite-to-device (DS2D)\acused{DS2D} communication has gained attention as a solution for reliable links between ground \ac{UE} and satellite receivers \cite{rappaport2025}. These scenarios involve \acp{UE} with limited transmission power and satellite receivers with constrained beamforming. Addressing similar challenges, \cite{kim2022} described a \ac{NB} \ac{IoT} \ac{UL} scenario with a ground \ac{UE} transmitting signals to a satellite receiver in the \ac{LEO}. The analysis therein showed that, despite adopting a transmit power of $\SI{23}{dBm}$ and covering a narrow bandwidth of $\SI{180}{\kilo\hertz}$ at $\SI{2}{\giga\hertz}$, robust channel coding yielding data rates as low as $\SI{4.6}{kbps}$ is needed to counter severe path loss and limited \ac{UE} antenna and array gains. When interference from terrestrial base stations, \ac{UE} devices, or other satellite \ac{UE} devices arises, countermeasures such as radio resource management or multiplexing are typically used \cite{ntontin2025}, trading-off data rate for robustness without adding further capabilities to the \ac{DS2D} \ac{UL}.

\begin{figure}[!t]
	\centering

    \psfrag{-A/2}[c][c]{\scriptsize $-L/2$}
    \psfrag{B/2}[c][c]{\scriptsize $L/2$}
	  \psfrag{L = vMT}[c][c]{\scriptsize $L=vMT$}
    \psfrag{m=0}[c][c]{\scriptsize $m=0$}
    \psfrag{m=M-1}[c][c]{\scriptsize $m=M-1$}
    \psfrag{R0}[c][c]{\scriptsize $R_0$}
    \psfrag{Rue}[c][c]{\scriptsize \textcolor{nokiapurple}{$R_\mathrm{UE}$}}
    \psfrag{Zue}[c][c]{\scriptsize \textcolor{nokiapurple}{$\theta_\mathrm{UE}$}}
    \psfrag{T}[c][c]{\scriptsize $\theta$}
    \psfrag{v}[c][c]{\scriptsize $v$}
    \psfrag{x}[c][c]{\scriptsize $x$}
    \psfrag{xUE}[c][c]{\scriptsize $x_\mathrm{UE}$}
    \psfrag{y}[c][c]{\scriptsize $y$}
	\includegraphics[height=6cm]{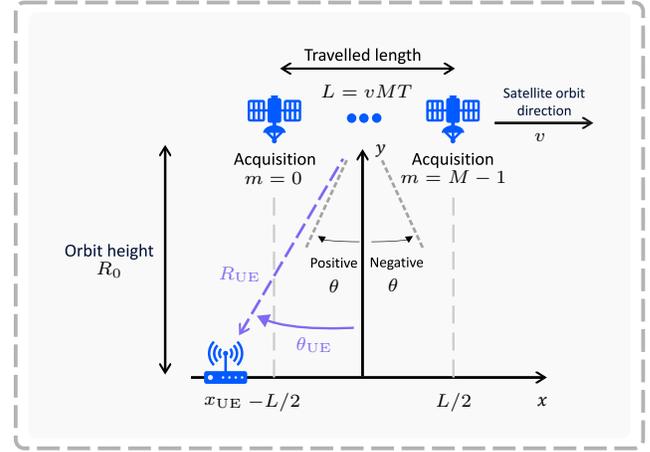}
	\captionsetup{justification=raggedright,labelsep=period,singlelinecheck=false}
	\caption{\ Satellite UL with LoS between a ground UE at \mbox{$x=x_\mathrm{UE}$} and \mbox{$y=0$} and a satellite at orbit height $R_0$ moving with orbital velocity $v$ from \mbox{$x=-L/2$} to \mbox{$x=L/2$} at \mbox{$y=R_0$}. From the middle point of the satellite's trajectory to the UE, a range $R_\mathrm{UE}$ and an azimuth angle $\theta_\mathrm{UE}$ are observed.}\label{fig:sysModel}
    \vspace{-0.385cm}
\end{figure}

To address these challenges, this article proposes a novel coherent \ac{SAC} system concept. As in Wi-Fi and future \ac{6G} standards, \ac{OFDM} is used as reference waveform. Then, \ac{SAR}-based processing is leveraged to make the experienced time-varying Doppler shift due to satellite movement constant, enabling coherent processing of consecutively acquired \ac{OFDM} symbol copies. Thanks to the satellite displacement, this forms a synthetic aperture providing coherent processing gain and enabling fine \ac{DoA} estimation. The latter facilitates estimating the direction of \ac{UE} devices and separating their signals in the angular domain, therefore also handling interference. The proposed concept differs from \ac{SAC} approaches in \ac{UWAC}, which rely on incoherent diversity combining \cite{he2017}. These approaches require long acquisition intervals for diversity, lack \ac{DoA} estimation, and do not address multi-user interference.

The proposed \ac{SAC} system concept can significantly advance \ac{DS2D} \ac{UL} capabilities, extending beyond currently envisioned use cases such as text messaging. 
It supports robust, low-data-rate communication and precise positioning in an \ac{ISAC} context, with \ac{DoA} estimation enabling fine cross-range resolutions. In addition, it supports massive \ac{IoT} applications in the \acused{NTN}non-terrestrial networks context, facilitating \ac{UL} communication between low-power or battery-limited \ac{UE} devices and receivers with well-defined trajectories, such as satellites.


\section{System Model}\label{sec:sys_model}

Let a satellite at an orbit height $R_0$ moving with orbital velocity $v$ along the $x$ axis as in Fig.~\ref{fig:sysModel} act as a receiver in an \ac{UL} scenario with bandwidth $B$ and carrier frequency~$f_\mathrm{c}$, where the transmitter is a static \ac{UE}. Furthermore, \ac{OFDM}-based communication is assumed, with a discrete-frequency domain frame \mbox{$\mathbf{S}\in\mathbb{C}^{N\times M}$} containing $M$ \ac{OFDM} symbols with $N$ subcarriers each being transmitted by the \ac{UE}. By performing \acp{IDFT} along the columns of $\mathbf{S}$, the discrete-time domain \ac{OFDM} frame \mbox{$\mathbf{s}\in\mathbb{C}^{N\times M}$} is generated. Next, a \ac{CP} of length $N_\mathrm{CP}$ is prepended to each of the $M$ \ac{OFDM} symbols in the columns of $\mathbf{s}$. The resulting discrete-time domain frame then undergoes \ac{P/S} conversion, \ac{D/A} conversion and further analog conditioning, yielding the complex baseband \ac{UE} transmit signal \mbox{$s(t)\in\mathbb{C}$} with transmit power $P_\mathrm{Tx}$. Additional relevant parameters of the \ac{OFDM} signal are its subcarrier spacing, symbol duration with \ac{CP}, and frame duration. The former is given by \mbox{$\Delta f =B/N $}. As for the latter two, they are defined as \mbox{$T = T_0 + T_\mathrm{CP}$}, where \mbox{$T_0=1/\Delta f=N/B$} is the \ac{OFDM} symbol duration without \ac{CP} and \mbox{$T_\mathrm{CP}=N_\mathrm{CP}/B$} is the \ac{CP} duration, and \mbox{$T_\mathrm{frame} = M\,T=M\,(N+N_\mathrm{CP})/B$}, respectively.

During the $M$ acquisitions of \ac{OFDM} symbols with duration $T$, the satellite travels a lengthmbox{$L = v\,M\,T$}. It is henceforth assumed that the length $L$ traveled by the satellite is between \mbox{$x=-L/2$} to \mbox{$x=L/2$} at the orbit height $R_0$, and that the \ac{UE} is located at \mbox{$x=x_\mathrm{UE}$} on the ground. The time-varying $x$ coordinate of the satellite is therefore \mbox{$x(t)=(-L/2) + vt$}. Consequently, the range between the \ac{UE} and the satellite is
\begin{equation}\label{eq:r_t}
    R(t) = \sqrt{{\left(x(t)-x_\mathrm{UE}\right)}^2+{R_0}^2}.
\end{equation}
After $s(t)$ being transmitted by the \ac{UE} with transmit antenna gain $G_\mathrm{Tx}$, and propagating through a \ac{LoS} path, the complex baseband representation of the satellite receive signal \mbox{$r(t)\in\mathbb{C}$} captured by a receive antenna of gain $G_\mathrm{Rx}$ is
\begin{equation}\label{eq:r_t}
    r(t) = \sqrt{G_\mathrm{Tx}\,G_\mathrm{Rx}}\,\alpha(t)\,s\left(t-\tau(t)\right)\e^{j\phi(t)} + \eta(t).
\end{equation}
In this equation, $a(t)$ is the time-varying attenuation that is influenced by path loss, atmospheric gas absorption losses, and ionospheric or tropospheric scintillation losses~\cite{3GPPTR38811}. Lastly, \mbox{$\eta(t)\in\mathbb{C}$} is the \ac{AWGN}. In \ac{NLoS} cases, building entry losses are also considered. Furthermore, $\tau(t)$ is the time-varying delay experienced by the \ac{UL} signal expressed as \mbox{$\tau(t)=R(t)/c_0$}, where $c_0$ is the speed of light in vacuum. \mbox{$\phi(t) = 2\pi f_\mathrm{c}\tau(t)=2\pi f_\mathrm{c}R(t)/c_0$} denotes the carrier phase, which is time varying due to the satellite movement. Based on \eqref{eq:r_t} and recalling that \mbox{$x(t)=(-L/2) + vt$}, it holds that
\begin{equation}\label{eq:phi_t_2}
    \phi(t) = 2\pi f_\mathrm{c}\frac{\sqrt{{\left(-\frac{L}{2} + v\,t - x_\mathrm{UE}\right)}^2+{R_0}^2}}{c_0}.
\end{equation}
For simplicity, it is henceforth assumed that the satellite is flying right above the \ac{UE} and that the traveled length $L$, and therefore also $v\,t$, is negligible compared to the satellite altitude $R_0$, i.e., \mbox{$L \ll R_0$}. Since \mbox{$L = v\,M\,T$}, \mbox{$L \ll R_0$} can be achieved with the transmission of a short number of \ac{OFDM} symbols $M$ for a given orbital velocity $v$ and \ac{OFDM} symbol duration $T$. Consequently, a time-invariant attenuation $\alpha$ and a time-invariant range \mbox{$R_\mathrm{UE}=\sqrt{{x_\mathrm{UE}}^2+{R_0}^2}$} and delay \mbox{$\tau=R_\mathrm{UE}/c_0$} are experienced. With larger $M$ and therefore larger apertures, however, range migration may happen and must be corrected \cite{moreira2013}. The time-varying behavior of the carrier $\phi(t)$ is kept, but for \mbox{$L \ll R_0$} and knowing that $c_0=\lambda~f_\mathrm{c}$, where $\lambda$ is the wavelength associated with the carrier frequency $f_\mathrm{c}$, it can be simplified as
\begin{equation}\label{eq:phi_t_2}
    \phi(t) = 2\pi \frac{\left[{\left(-\frac{L}{2} + v\,t - x_\mathrm{UE}\right)}^2/\left(2R_0\right)\right]+R_0}{\lambda}.
\end{equation}
Consequently, \eqref{eq:r_t} becomes
\begin{align}\label{eq:r_t_2}
    r(t) &= \sqrt{G_\mathrm{Tx}\,G_\mathrm{Rx}}\,\alpha\,s\left(t-\frac{R_\mathrm{UE}}{c_0}\right)\nonumber\\
    &\phantom{=}\cdot~\e^{\im2\pi\frac{\left[{\left(-\frac{L}{2} + v\,t - x_\mathrm{UE}\right)}^2/\left(2R_0\right)\right]+R_0}{\lambda}} + \eta(t).
\end{align}
Based on the expression of $\phi(t)$ in \eqref{eq:phi_t_2}, the experienced time-varying Doppler shift $f_\mathrm{D}(t)$ by $r(t)$ can be calculated as
\begin{equation}\label{eq:f_d_t}
    f_\mathrm{D,UE}(t) = \frac{1}{2\pi}\,\frac{d}{dt}\,\phi(t) = -\frac{\left(v\,\frac{L}{2} + x_\mathrm{UE}\right)}{R_0\,\lambda} + \frac{v^2\,t}{R_0\,\lambda}.
\end{equation}

\section{Coherent Synthetic Aperture Combining}\label{sec:syn_apt}

In the considered \ac{DS2D} \ac{UL} scenario, each \ac{UE} transmits $M$ copies of the same \ac{OFDM} symbol. While this results in a data rate reduction by a factor of $M$, it allows exploiting the time-varying Doppler shift experienced due to the satellite movement to synthesize a virtual aperture for the satellite receiver. Although not required, ideal synchronization is assumed, and the processing in the \ac{SAC} system is described as follows.

\subsection{Azimuth compression}\label{subsec:az_comp}

After analog conditioning, \ac{A/D} conversion, and \ac{S/P} conversion, the \ac{CP} is removed from all $M$ \ac{OFDM} symbols. The resulting receive \ac{OFDM} frame in the discrete-time domain is denoted by \mbox{$\mathbf{r}\in\mathbb{C}^{N\times M}$}. The $n\mathrm{th}$ sample, \mbox{$n\in\{0, 1, \dots, N-1\}$}, of the $m\mathrm{th}$ \ac{OFDM} symbol, \mbox{$m\in\{0, 1, \dots, M-1\}$}, which is located in row $n$ and column $m$ of $\mathbf{r}$, is given by
\begin{equation}\label{eq:r_n_m}
    r_{n,m} = r(t)\rvert_{t=mT+T_\mathrm{CP}+n/B}.
\end{equation}
Next, to eventually make the experienced time-varying Doppler shift $f_\mathrm{D,UE}(t)$ constant, azimuth compression is performed along the \ac{OFDM} symbols in $\mathbf{r}$. Still assuming pure \ac{LoS} between \ac{UE} and satellite, this is done by multiplying every sample $r_{n,m}$ by \mbox{$h_{\mathrm{az},n,m} = \e^{-\im\phi_{\mathrm{az},n,m}}$}, in which
\begin{equation}
    \phi_{\mathrm{az},n,m} = 2\pi \frac{\left\{{\left[-\frac{L}{2} + v\,\left(mT+T_\mathrm{CP}+\frac{n}{B}\right)\right]}^2/\left(2R_0\right)\right\}+R_0}{\lambda}
\end{equation}
is the phase response according to \eqref{eq:phi_t_2} for an \ac{UE} at \mbox{$x_\mathrm{UE}=\SI{0}{\meter}$} on the ground \cite{moreira2013} evaluated at the \mbox{$m\mathrm{th}$} \ac{OFDM} symbol, which starts at \mbox{$t=mT+T_\mathrm{CP}$}. This operation results in the matrix \mbox{$\mathbf{r}_\mathrm{az}\in\mathbb{C}^{N\times M}$}, whose element \mbox{$r_{\mathrm{az},n,m}=r_{n,m}\,h_{\mathrm{az},n,m}$} at its \mbox{$n\mathrm{th}$} row and \mbox{$m\mathrm{th}$} column is equal to
\begin{align}\label{eq:r_n_m_az_1}
    r_{\mathrm{az},n,m} &= \sqrt{G_\mathrm{Tx}\,G_\mathrm{Rx}}\,\alpha\,s\left(\left(mT+T_\mathrm{CP}+\frac{n}{B}\right)-\frac{R_\mathrm{UE}}{c_0}\right)\nonumber\\
    &\phantom{=}\,\cdot~\e^{\im\frac{2\pi}{R_0\,\lambda}\left[\frac{{x_\mathrm{UE}}^2+L\,x_\mathrm{UE}}{2}-v\,x_\mathrm{UE}\,\left(mT+T_\mathrm{CP}+\frac{n}{B}\right)\right]}\nonumber\\
    &\phantom{=} +~\eta_{n,m}\,\e^{-\im\,\phi_{\mathrm{az},n,m}}.
\end{align}
In this equation, \mbox{$\eta_{n,m}\in\mathbb{C}$} is a Gaussian random variable with mean \mbox{$\mathbb{E}\{\eta_{n,m}\}=0$} and variance \mbox{$\mathbb{E}\{\lvert\eta_{n,m}\rvert^2\}=\sigma_\eta^2$} that represents the effect of the \ac{AWGN} $\eta(t)$ in the \mbox{$n\mathrm{th}$} row and \mbox{$m\mathrm{th}$} column of $\mathbf{r}_\mathrm{az}$. The resulting noise $\eta_{n,m}\,\e^{-\im\,\phi_{\mathrm{az},n,m}}$ in this equation is henceforth denoted as \mbox{$\eta_{\mathrm{az},n,m}\in\mathbb{C}$}, and it has the same mean and variance as $\eta_{n,m}$, i.e., \mbox{$\mathbb{E}\{\eta_{\mathrm{az},n,m}\}=0$} and \mbox{$\mathbb{E}\{\lvert\eta_{\mathrm{az},n,m}\rvert^2\}=\,\sigma_\eta^2$}. Furthermore, as all $M$ transmit \ac{OFDM} symbols are equal as assumed at the beginning of this section, \eqref{eq:r_n_m_az_1} can be simplified 
and rearranged as
\begin{align}\label{eq:r_n_m_az_3}
    r_{\mathrm{az},n,m} &= \sqrt{G_\mathrm{Tx}\,G_\mathrm{Rx}}\,\alpha\,s\left(\left(T_\mathrm{CP}+\frac{n}{B}\right)-\frac{R_\mathrm{UE}}{c_0}\right)\nonumber\\
    &\phantom{=} \cdot~\e^{\im\frac{2\pi}{R_0\,\lambda}\left(\frac{{x_\mathrm{UE}}^2+L\,x_\mathrm{UE}}{2}\right)}\,\e^{\im\,2\,\pi\left(\frac{-v\,x_\mathrm{UE}}{R_0\,\lambda}\right)\,\left(mT+T_\mathrm{CP}+\frac{n}{B}\right)}\nonumber\\
    &\phantom{=} +~\eta_{\mathrm{az},n,m}.
\end{align}
An analysis of this equation reveals that a Doppler shift $f_\mathrm{D,UE}$ is experienced, which is expressed as
\begin{equation}\label{eq:f_d_ideal}
    f_\mathrm{D,UE} = -\frac{v\,x_\mathrm{UE}}{R_0\,\lambda}.
\end{equation}
As previously mentioned, the ultimately experienced Doppler shift $f_\mathrm{D,UE}$ results from the time-varying one $f_\mathrm{D,UE}(t)$, which is made constant by the performed azimuth compression.

\subsection{Doppler shift estimation}\label{subsec:f_d_est}

By performing \acp{DFT} along the columns of $\mathbf{r}_\mathrm{az}$, a matrix \mbox{$\mathbf{y}_\mathrm{az}\in\mathbb{C}^{N\times M}$} is obtained, whose element at its \mbox{$n\mathrm{th}$} row and \mbox{$l\mathrm{th}$} column, \mbox{$l\in\{-M/2, -M/2+1, \dots, M/2-1\}$}, is given by
\begin{align}\label{eq:y_n_k_az_2}
    y_{\mathrm{az},n,l} &= \frac{1}{\sqrt{M}}\,\sqrt{G_\mathrm{Tx}\,G_\mathrm{Rx}}\,\alpha\,s\left(\left(T_\mathrm{CP}+\frac{n}{B}\right)-\frac{R_\mathrm{UE}}{c_0}\right)\nonumber\\
    &\phantom{=} \cdot~\e^{\im\frac{2\pi}{R_0\,\lambda}\left(\frac{{x_\mathrm{UE}}^2+L\,x_\mathrm{UE}}{2}\right)}\,\e^{\im\,2\,\pi\left(\frac{-v\,x_\mathrm{UE}}{R_0\,\lambda}\right)\,\left(T_\mathrm{CP}+\frac{n}{B}\right)}\nonumber\\
    &\phantom{=} \cdot~\sum_{m=0}^{M-1} \e^{-\im\frac{2\pi\,m}{M}\left(l+\frac{v\,x_\mathrm{UE}\,T\,M}{R_0\,\lambda}\right)} +\eta'_{\mathrm{az},n,l}.
\end{align}
In this equation, $\eta'_{\mathrm{az},n,l}$ is the \ac{DFT} of $\eta_{\mathrm{az},n,m}$, which has the same mean \mbox{$\mathbb{E}\{\eta'_{\mathrm{az},n,m}\}=0$} and variance \mbox{$\mathbb{E}\{\lvert\eta'_{\mathrm{az},n,m}\rvert^2\}=\sigma_\eta^2$} since the \ac{DFT} was normalized.

Next, the Euclidean norm of the columns of $\mathbf{y}_\mathrm{az}$ is taken to produce the vector \mbox{$\mathbf{y}_\mathrm{az,norm}\in\mathbb{C}^{1\times M}$}. Based on \eqref{eq:y_n_k_az_2}, a maximum occurs at the \mbox{$\widehat{l}_\mathrm{UE}\mathrm{th}$} column of $\mathbf{y}_\mathrm{az,norm}$, \mbox{$\widehat{l}_\mathrm{UE}\in\{-M/2, -M/2+1, \dots, M/2-1\}$}, such that
\begin{equation}\label{eq:k_max}
    \widehat{l}_\mathrm{UE} = -\frac{v\,x_\mathrm{UE}\,T\,M}{R_0\,\lambda}.
\end{equation}
The Doppler shift resolution $\Delta f_\mathrm{D}$ for \ac{DFT}-based estimation in an \ac{OFDM}-based system is given by
\begin{equation}\label{eq:delta_f_d}
    \Delta f_\mathrm{D} = \frac{1}{T_\mathrm{frame}} = \frac{1}{M\,T}.
\end{equation}
Knowing that $\Delta f_\mathrm{D}$ is equal to the Doppler shift bin width, the sample estimate $\widehat{l}_\mathrm{UE}$ calculated according to \eqref{eq:k_max} is associated with a Doppler shift estimate \mbox{$\widehat{f}_\mathrm{D,UE}=\widehat{l}_\mathrm{UE}\,\Delta f_\mathrm{D}$}, which for high \ac{SNR} is approximately equal to $f_\mathrm{D}$.

\subsection{Azimuth estimation}\label{subsec:az_est}

To convert the Doppler shift estimate $\widehat{f}_\mathrm{D,UE}$ into an azimuth estimate, the angle $\theta_\mathrm{UE}$ in radians of the \ac{UE} w.r.t. to the point in the center of the satellite trajectory during the transmission of the $M$ \ac{OFDM} symbols, i.e., \mbox{$x=\SI{0}{\meter}$} at the orbit height $R_0$, can be calculated based on Fig.~\ref{fig:sysModel} as
\begin{equation}\label{eq:theta_est_0}
    \theta_\mathrm{UE} = \arcsin\left(\frac{x_\mathrm{UE}}{\sqrt{{x_\mathrm{UE}}^2+{R_0}^2}}\right) \stackrel{\lvert x_\mathrm{UE}\vert\ll R_0 }{=} \frac{x_\mathrm{UE}}{R_0}.
\end{equation}
Combining \eqref{eq:f_d_ideal} and \eqref{eq:theta_est_0} allows expressing the azimuth estimate $\widehat{\theta}_\mathrm{UE}$ for the \ac{UE} as
\begin{equation}\label{eq:theta_f_d}
    \widehat{\theta}_\mathrm{UE} = - \frac{\widehat{f}_\mathrm{D,UE}\,\lambda}{v}.
\end{equation}
The expression in \eqref{eq:theta_f_d} shows that, by exploiting the experienced Doppler shift by the movement of the satellite receiver, a virtual array is synthesized, and angular estimation is enabled.

\subsection{Azimuth resolution and ambiguity}\label{subsec:az_res_amb}

Based on \eqref{eq:delta_f_d} and \eqref{eq:theta_f_d}, the azimuth resolution associated with the virtual array is given by
\begin{equation}\label{eq:delta_theta}
    \Delta\theta = \frac{\lambda}{v\,M\,T} = \frac{\lambda}{L}.
\end{equation}
This equation allows interpreting the traveled length $L$ by the satellite as the synthetic aperture length \cite{moreira2013}.

A further relevant parameter is the maximum unambiguous azimuth. It is derived from the maximum unambiguous Doppler shift for \ac{DFT}-based estimation in an \ac{OFDM}-based systems and expressed as
\begin{equation}\label{eq:theta_ua}
    \theta_\mathrm{max,ua} = \pm\frac{\lambda}{2\,v\,T} = \pm\frac{M}{2}\Delta\theta.
\end{equation}
It is worth highlighting that residual \ac{CFO} may cause angular estimates to be biased or folded around the unambiguous azimuth range. Even so, \ac{UE} devices with angular separation of at least $\Delta\theta$ can still be resolved.

\subsection{Further communication processing}\label{subsec:comm_proc}

The processing on the matrix $\mathbf{r}_\mathrm{az}$ to extract the modulation symbols transmitted by the \ac{UE} is described as follows. First, a Doppler shift correction is performed by multiplying $\mathbf{r}_\mathrm{az}$ by the steering vector \mbox{$\mathbf{b}(\mathrm{\theta})\in\mathbb{C}^{M\times 1}$} evaluated for the obtained azimuth estimate $\widehat{\theta}_\mathrm{UE}$. Considering that the distance \mbox{$d=v\,T$} traveled by the satellite between the transmission of consecutive \ac{OFDM} symbols is equal to the antenna element spacing in the synthesized array with synthetic aperture and that \mbox{$\sin\left(\theta\right)\approx\theta$} for small $\theta$, the element at the \mbox{$m\mathrm{th}$} row of the receive steering vector is defined as \mbox{$b_m(\theta) = (1/\sqrt{M})\,e^{j \frac{2\pi}{\lambda}\,m\,v\,T\,\theta}$}. Performing receive beamsteering of $\mathbf{r}_\mathrm{az}$ for the azimuth estimate $\widehat{\theta}_\mathrm{UE}$ yields \mbox{$\mathbf{r}_{\widehat{\theta}_\mathrm{UE}}\in\mathbb{C}^{N\times 1}|\mathbf{r}_{\widehat{\theta}}=\mathbf{r}_\mathrm{az}\mathbf{b}(\widehat{\theta}_\mathrm{UE})$}. 
Next, $\mathbf{r}_{\widehat{\theta}_\mathrm{UE}}$ undergoes Doppler shift correction by multiplying each \mbox{$n\mathrm{th}$} element with \mbox{$\e^{-\im\,2\pi\,\widehat{f}_\mathrm{D,UE}\,\frac{n}{B}}$}. Based on \eqref{eq:f_d_ideal} and assuming an ideal Doppler shift estimate \mbox{$\widehat{f}_\mathrm{D,UE}=f_\mathrm{D,UE}$}, this complex exponential is equal to \mbox{$\e^{\im\,\frac{2\pi}{R_0\,\lambda}\,v\,x_\mathrm{UE}\,\frac{n}{B}}$}. This operation results in the vector \mbox{$\mathbf{r}_{\widehat{\theta}_\mathrm{UE},D}\in\mathbb{C}^{N\times 1}$},
which after undergoing \ac{DFT} results in \mbox{$\mathbf{R}_{\widehat{\theta}_\mathrm{UE},\mathrm{D}}\in\mathbb{C}^{N\times 1}$}. The \mbox{$k\mathrm{th}$} element of $\mathbf{R}_{\widehat{\theta}_\mathrm{UE},\mathrm{D}}$ is given by
\begin{align}\label{eq:R_theta_d_k}
    &R_{\widehat{\theta}_\mathrm{UE},\mathrm{D},k} = \sqrt{G_\mathrm{p,az}}\,\left[\sqrt{G_\mathrm{Tx}\,G_\mathrm{Rx}}\,\alpha\,S_k\,\e^{-\im\,\frac{2\pi\,k}{N}\left(\frac{R_\mathrm{UE}\,B}{c_0}\right)}\right.\nonumber\\
    &\left.\cdot~\e^{\im\frac{2\pi}{R_0\,\lambda}\left(\frac{{x_\mathrm{UE}}^2+L\,x_\mathrm{UE}}{2}\right)}\,\e^{\im\,2\,\pi\left(\frac{-v\,x_\mathrm{UE}}{R_0\,\lambda}\right)\,T_\mathrm{CP}}\right]+ \eta_{\widehat{\theta}_\mathrm{UE},\mathrm{D},k},
\end{align}
where $S_k$ is the transmit modulation symbol at the $k\mathrm{th}$ subcarrier of one of the $M$ repeated \ac{OFDM} symbols and $\eta_{\widehat{\theta}_\mathrm{UE},\mathrm{D},k}$ is the noise component which has the same mean \mbox{$\mathbb{E}\{\eta_{\widehat{\theta}_\mathrm{UE},\mathrm{D},k}\}=0$} and variance \mbox{$\mathbb{E}\{\lvert\eta_{\widehat{\theta}_\mathrm{UE},\mathrm{D},k}\rvert^2\}=\sigma_\eta^2$} as the one in $\mathbf{r}_\mathrm{az}$ since beamsteering and \ac{DFT} were normalized.

Assuming ideal \ac{CSI} is available, 
\ac{ZF} equalization can be performed on $\mathbf{R}_{\widehat{\theta}_\mathrm{UE},\mathrm{D}}$ to yield \mbox{$\mathbf{R}\in\mathbb{C}^{N\times 1}|\mathbf{R}=\mathbf{R}_{\widehat{\theta}_\mathrm{UE},\mathrm{D}}\odot\mathbf{G}$}, where $\odot$ is the Hadamard product and \mbox{$\mathbf{G}\in\mathbb{C}^{N\times 1}$} is the \ac{ZF} equalizer. For a pure \ac{LoS} channel, the \mbox{$k\mathrm{th}$} element of $\mathbf{R}$ is given by
\begin{equation}\label{eq:R_k}
    R_k = S_k + \eta'_{\widehat{\theta}_\mathrm{UE},\mathrm{D},k},
\end{equation}
where $\eta'_{\widehat{\theta}_\mathrm{UE},\mathrm{D},k}$ is the modified noise by the \ac{ZF} equalizer, which, however, has unchanged mean \mbox{$\mathbb{E}\{\eta'_{\widehat{\theta}_\mathrm{UE},\mathrm{D},k}\}=0$} and variance \mbox{$\mathbb{E}\{\lvert\eta'_{\widehat{\theta}_\mathrm{UE},\mathrm{D},k}\rvert^2\}=\sigma_\eta^2$}. The noise variance is given by \mbox{$\sigma_\eta^2 = N\,P_\eta$}, where $P_\eta$ is the \ac{AWGN} power given by \mbox{$P_\eta = k_\mathrm{B}\,\Delta f\,T_\mathrm{therm}\,\mathrm{NF}$}. In this equation, $k_\mathrm{B}$ is the Boltzmann constant, $T_\mathrm{therm}$ is the standard room temperature in Kelvin, and $\mathrm{NF}$ is the overall receiver noise figure in linear scale. Recalling that \mbox{$B=N\,\Delta f$} and assuming uniform power allocation among the $N$ subcarriers, the \ac{SNR} associated with the estimate of the transmit modulation symbol $S_k$ with \eqref{eq:R_k} is
\begin{equation}\label{eq:SNR_final}
    \mathrm{SNR} = \frac{P_\mathrm{Tx}\,G_\mathrm{Tx}\,G_\mathrm{Rx}\,G_\mathrm{p,az}\,\alpha^2}{k_\mathrm{B}\,B\,T_\mathrm{therm}\,\mathrm{NF}}.
\end{equation}


\section{Numerical Analysis}\label{sec:analysis}

\begin{table}[!t]
    \centering
	\captionsetup{width=43pc,justification=centering,labelsep=newline}
	\caption{\textsc{Synthetic Aperture Communication System Parameter\\ Requirements for Sub-Kilometer Cross-Range Resolution}}\label{tab:sac_param_subkm}
    \begin{tabular}{|c|c|c|c|}
        \hhline{|====|}
        $\bm{\Delta f}$ & $\bm{M}$ & $\bm{G_\mathrm{p,az}}$ & $\bm{\mathcal{R}_\mathrm{b}}$ \\ \hhline{|====|}
        \multirow{2}{*}{$\SI{15}{\kilo\hertz}$} & \multirow{2}{*}{$\geq93$} & \multirow{2}{*}{$\geq\SI{19.68}{dB}$} & $\leq\SI{45.21}{\kilo\bit/\second}$ ($B=\SI{4.50}{\mega\hertz}$) \\
         &  &  & $\leq\SI{488.23}{\kilo\bit/\second}$ ($B=\SI{48.60}{\mega\hertz}$) \\ \hline
        \multirow{2}{*}{$\SI{30}{\kilo\hertz}$} & \multirow{2}{*}{$\geq185$} & \multirow{2}{*}{$\geq\SI{22.67}{dB}$} & $\leq\SI{20.00}{\kilo\bit/\second}$ ($B=\SI{3.96}{\mega\hertz}$) \\
         &  &  & $\leq\SI{496.40}{\kilo\bit/\second}$ ($B=\SI{98.28}{\mega\hertz}$) \\ \hline
        \multirow{2}{*}{$\SI{60}{\kilo\hertz}$} & \multirow{2}{*}{$\geq369$} & \multirow{2}{*}{$\geq\SI{25.67}{dB}$} & $\leq\SI{20.06}{\kilo\bit/\second}$ ($B=\SI{7.92}{\mega\hertz}$) \\
         &  &  & $\leq\SI{246.14}{\kilo\bit/\second}$ ($B=\SI{97.20}{\mega\hertz}$) \\ \hhline{|====|}
    \end{tabular}
    \vspace{-0.385cm}
\end{table}

In this section, the proposed coherent \ac{SAC} system concept is verified via simulations. The  satellite receiver is assumed to be in the \ac{LEO} with orbit height \mbox{$R_0=\SI{600}{\kilo\meter}$} and orbital velocity \mbox{$v=\SI{7.82}{\kilo\meter/\second}$}. The channel model from \cite{3GPPTR38811} for a pure \ac{LoS} case at \mbox{$f_\mathrm{c}=\SI{3.5}{\giga\hertz}$} is considered, with losses including a path loss of $\SI{158.89}{dB}$, atmospheric absorption losses of $\SI{0.12}{dB}$, and scintillation losses of $\SI{4.39}{dB}$ between \ac{UE} transmitters at \mbox{$y=\SI{0}{\meter}$} and the satellite receiver orbiting at \mbox{$y=\SI{600}{\kilo\meter}$}, with negligible deviations during the satellite displacement along the $x$ axis for the considered scenario. Combined, these yield a total propagation loss of $\SI{163.40}{dB}$.

Next, it is assumed that each \ac{UE} was equipped with a $\lambda/2$-spaced \ac{URA} with fixed array aperture of $\SI[parse-numbers=false]{8.56\times8.56}{\cm^2}$, a gain of \mbox{$G_\mathrm{Tx}=\SI{11.72}{dBi}$}, and a \ac{HPBW} of $\SI{50.6}{\degree}$ \cite{cui2025}. For the satellite receiver, the gain of \mbox{$G_\mathrm{Rx}=\SI{30}{dBi}$ and the \ac{HPBW} of $\SI{4.41}{\degree}$} from Table \mbox{6.1.1.1-1} of the 3GPP TS 38.821 \cite{3GPPTR38821}, as well as a noise figure \mbox{$\mathrm{NF}=\SI{4}{dB}$} \cite{shahid2024} are assumed.

To further alleviate transmit power requirements of the \ac{SAC} system in addition to the azimuth processing gain $G_\mathrm{p,az}$ and ensure tolerable communication error rates, a \ac{5G NR} polar code with rate \mbox{$\mathcal{R}_\mathrm{c}=2/3$} is henceforth adopted. In addition, $25\%$ out of the total of $N$ subcarriers are used as pilots in a regularly-spaced comb. After coherent synthetic aperture combining, these pilots are extracted and used for channel estimation. The obtained channel frequency response is then interpolated to cover all $N$ subcarriers and enable subsequent equalization.

\subsection{Synthetic aperture communication system parameters}\label{subsec:system_params}

For sub-kilometer cross-range resolution, a synthetic aperture length \mbox{$L>\SI{51.39}{\meter}$} is required based on \eqref{eq:delta_theta}. For subcarrier spacings in \ac{FR1} numerologies \mbox{$\mu=0$} to \mbox{$\mu=2$} of \ac{5G NR}, this results in the requirements in Table~\ref{tab:sac_param_subkm}. For the performed simulations, standard-compliant normal \ac{CP} durations were considered \cite{3GPPTS38211}. The numbers in Table~\ref{tab:sac_param_subkm} show that such fine resolution is associated with azimuth processing gains that significantly alleviate the link budget. This comes, however, at the cost of reduction of the net data rate $\bm{\mathcal{R}_\mathrm{b}}$ by a factor of $M$ compared to the case where no synthetic aperture is synthesized and different \ac{OFDM} symbols are transmitted. Note that, although code spreading can achieve a similar trade-off between \ac{SNR} improvement and data rate, it lacks the \ac{DoA} estimation and angular resolution and user multiplexing capabilities inherent to the introduced \ac{SAC} system concept.

\begin{figure}[!t]
	\centering
	\subfloat[ ]{					
    	\psfrag{-5}[c][c]{\scriptsize -$5$}
    	\psfrag{-2.5}[c][c]{\scriptsize -$2.5$}
    	\psfrag{0}[c][c]{\scriptsize $0$}
    	\psfrag{2.5}[c][c]{\scriptsize $2.5$}
    	  \psfrag{5}[c][c]{\scriptsize $5$}

        \psfrag{2}[c][c]{\scriptsize $2$}
    	\psfrag{0}[c][c]{\scriptsize $0$}
    	\psfrag{-2}[c][c]{\scriptsize -$2$}
    	\psfrag{-4}[c][c]{\scriptsize -$4$}
    	\psfrag{-6}[c][c]{\scriptsize -$6$}
        \psfrag{-8}[c][c]{\scriptsize -$8$}
        \psfrag{-10}[c][c]{\scriptsize -$10$}
    		
    	\psfrag{x (km)}[c][c]{\footnotesize $x$ (km)}
    	\psfrag{Norm. mag. (dB)}[c][c]{\footnotesize Norm. mag. (dB)}

        \includegraphics[height=4cm]{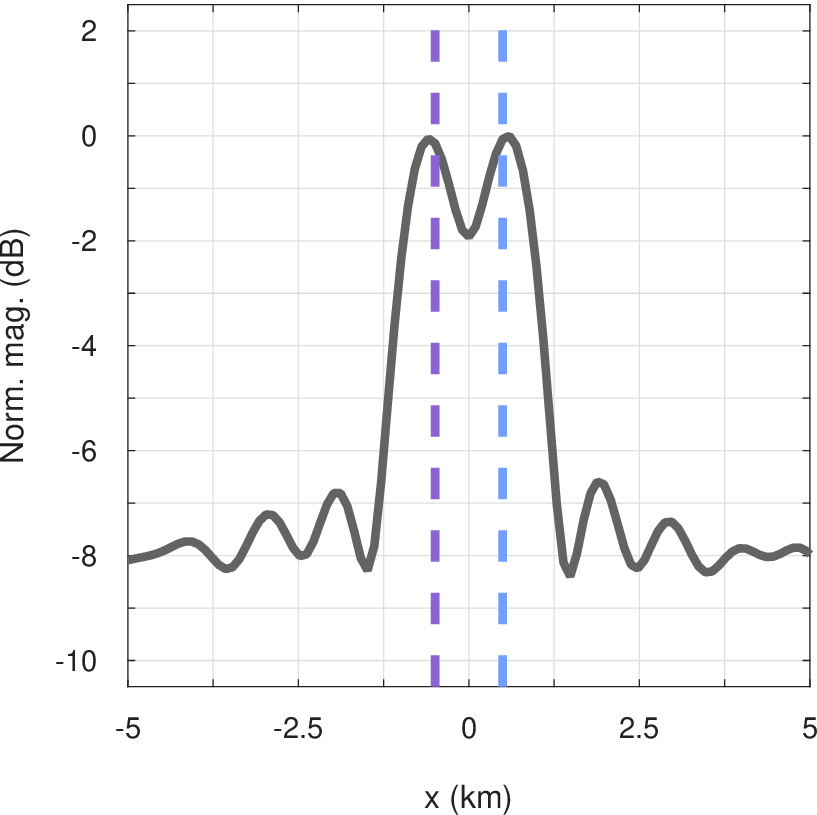}\label{fig:mu_az_est_a}
    }\hspace{0.2cm}
    \subfloat[ ]{	
        \psfrag{-5}[c][c]{\scriptsize -$5$}
    	\psfrag{-2.5}[c][c]{\scriptsize -$2.5$}
    	\psfrag{0}[c][c]{\scriptsize $0$}
    	\psfrag{2.5}[c][c]{\scriptsize $2.5$}
    	  \psfrag{5}[c][c]{\scriptsize $5$}
    			
    	\psfrag{2}[c][c]{\scriptsize $2$}
        \psfrag{0}[c][c]{\scriptsize $0$}
    	\psfrag{-2}[c][c]{\scriptsize -$2$}
    	\psfrag{-4}[c][c]{\scriptsize -$4$}
    	\psfrag{-6}[c][c]{\scriptsize -$6$}
        \psfrag{-8}[c][c]{\scriptsize -$8$}
        \psfrag{-10}[c][c]{\scriptsize -$10$}
    		
    	\psfrag{x (km)}[c][c]{\footnotesize $x$ (km)}
    	\psfrag{Norm. mag. (dB)}[c][c]{\footnotesize Norm. mag. (dB)}
    			
    	\includegraphics[height=4cm]{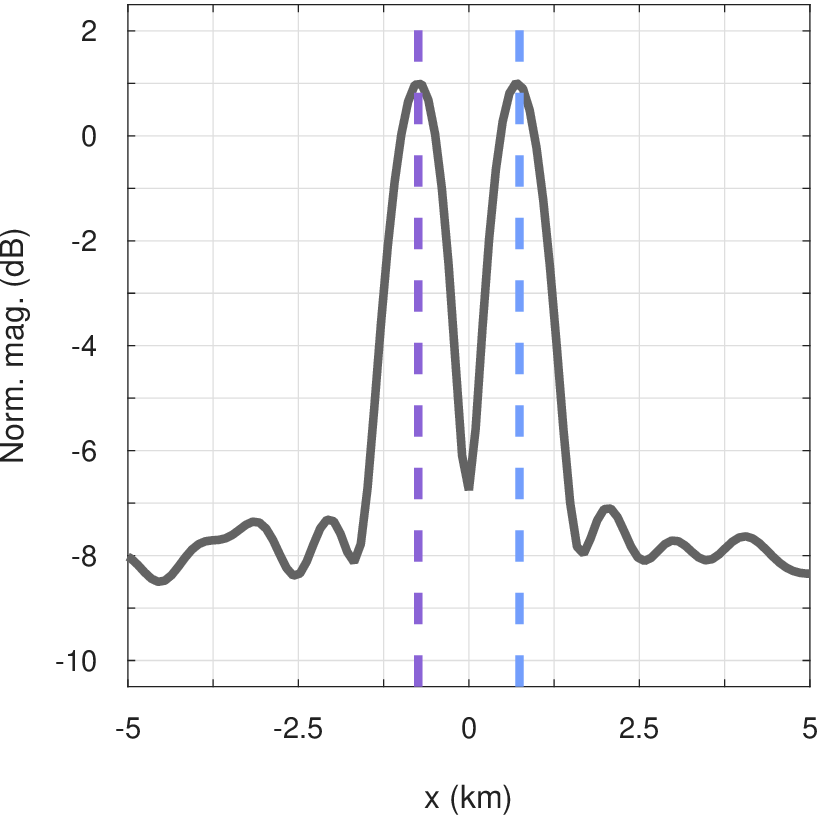}\label{fig:mu_az_est_b}
    }
	\captionsetup{justification=raggedright,labelsep=period,singlelinecheck=false}
	\caption{\ Azimuth profile as a function of the $x$ coordinate ({\color[rgb]{0.3922,0.3922,0.3922}\rule[0.5ex]{1em}{1.5pt}}). For (a), two perfectly resolved \ac{UE} devices are considered,   $x_\mathrm{UE,1}=\SI{-495.33}{\meter}$ \mbox{({\color[rgb]{0.5412,0.3922,0.8392}\rule[0.5ex]{0.4em}{1.5pt}\hspace{0.2em}\rule[0.5ex]{0.4em}{1.5pt}})} and $x_\mathrm{UE,2}=\SI{495.33}{\meter}$ \mbox{({\color[rgb]{0.4471,0.6196,0.9922}\rule[0.5ex]{0.4em}{1.5pt}\hspace{0.2em}\rule[0.5ex]{0.4em}{1.5pt}})}. In (b), two resolved \ac{UE} devices, $x_\mathrm{UE,1}=\SI{-742.99}{\meter}$ ({\color[rgb]{0.5412,0.3922,0.8392}\rule[0.5ex]{0.4em}{1.5pt}\hspace{0.2em}\rule[0.5ex]{0.4em}{1.5pt}}) and  $x_\mathrm{UE,2}=\SI{742.99}{\meter}$ ({\color[rgb]{0.4471,0.6196,0.9922}\rule[0.5ex]{0.4em}{1.5pt}\hspace{0.2em}\rule[0.5ex]{0.4em}{1.5pt}}), are considered. All magnitudes are normalized w.r.t. to the highest peak in (a).}\label{fig:az_profiles}
    \vspace{-0.385cm}
\end{figure}

\subsection{Communication performance in multi-user scenario}\label{subsec:mu_perf}

Next, a scenario with 2 \ac{UE} devices, labeled as \ac{UE} \#1 and \ac{UE} \#2, performing \ac{UL} transmission with full time and frequency overlap is assumed. All previously adopted parameters are kept except for subcarrier spacing, bandwidth, and number of acquired \ac{OFDM} symbol copies, which are set to \mbox{$\Delta f=\SI{15}{\kilo\hertz}$}, \mbox{$B=\SI{4.5}{\mega\hertz}$}, and \mbox{$M=93$}, respectively. Consequently, a synthetic aperture length \mbox{$L=\SI{51.88}{\meter}$} is obtained, which is associated with an azimuth resolution of \mbox{$\Delta\theta=\SI[parse-numbers=false]{\left(94.60\cdot10^{-3}\right)}{\degree}$} and a maximum unambiguous azimuth of \mbox{$\theta_\mathrm{max,ua}=\pm\SI{4.40}{\degree}$}. In addition, these values correspond to a cross range resolution of \mbox{$R_0\,\Delta\theta=\SI{990.65}{\meter}$} and a maximum unambiguous cross range of \mbox{$R_0\,\theta_\mathrm{max,ua}=\pm\SI{46.07}{\kilo\meter}$}.

Fig.~\ref{fig:az_profiles} shows the result of the procedure to obtain the azimuth estimates $\widehat{\theta}_\mathrm{UE}$ and consequently $x$ coordinate estimates of \ac{UE} devices described in Sections~\ref{subsec:az_comp} to \ref{subsec:f_d_est} for a transmit power \mbox{$P_\mathrm{Tx}=\SI{-10}{dBm}$} and two scenarios. These are, namely, $x_\mathrm{UE,1}=\SI{-495.33}{\meter}$ for \ac{UE} \#1 and $x_\mathrm{UE,2}=\SI{495.33}{\meter}$ for \ac{UE} \#2 in Fig.~\ref{fig:az_profiles}(a), and $x_\mathrm{UE,1}=\SI{-742.99}{\meter}$ for \ac{UE} \#1 and $x_\mathrm{UE,2}=\SI{742.99}{\meter}$ for \ac{UE} \#2 in Fig.~\ref{fig:az_profiles}(b). In the first, the \ac{UE} devices are separated by $\SI{990.65}{\meter}$, an integer multiple of the cross-range resolution, which allows full separability of their signals. In the second scenario, however, the cross-range difference between \ac{UE} devices is $\SI{1.49}{\kilo\meter}$, which is equal to \mbox{$1.5\cdot R_0\,\theta_\mathrm{max,ua}$}. This distance is not an integer multiple of the resolution and leads to the maximum azimuth sidelobe level of around $\SI{-13.26}{dB}$ w.r.t. the main peak from one \ac{UE} at the position of the other one, which is observed in the form of higher normalized magnitude at the \ac{UE} cross ranges. Consequently, maximum mutual interference between resolved \ac{UE} devices in the angular domain is experienced during the beamsteering described in Section~\ref{subsec:comm_proc}.

\begin{figure}[!t]
	\centering
	\psfrag{-15}[c][c]{\scriptsize -$15$}
    \psfrag{-10}[c][c]{\scriptsize -$10$}
    \psfrag{-5}[c][c]{\scriptsize -$5$}
    \psfrag{0}[c][c]{\scriptsize $0$}
    \psfrag{5}[c][c]{\scriptsize $5$}
    \psfrag{10}[c][c]{\scriptsize $10$}

    \psfrag{~19.82 dB}[c][c]{\scriptsize $\SI[parse-numbers=false]{\sim20}{dB}$}
    			
   	\psfrag{0.1}[c][c]{\scriptsize $0.1$}
   	\psfrag{1}[c][c]{\scriptsize $1$}
    		
	\psfrag{Transmit power (dBm)}[c][c]{\footnotesize $P_\mathrm{Tx}$ (dBm)}
    	\psfrag{BLER}[c][c]{\footnotesize BLER}
    			
    \includegraphics[height=4.15cm]{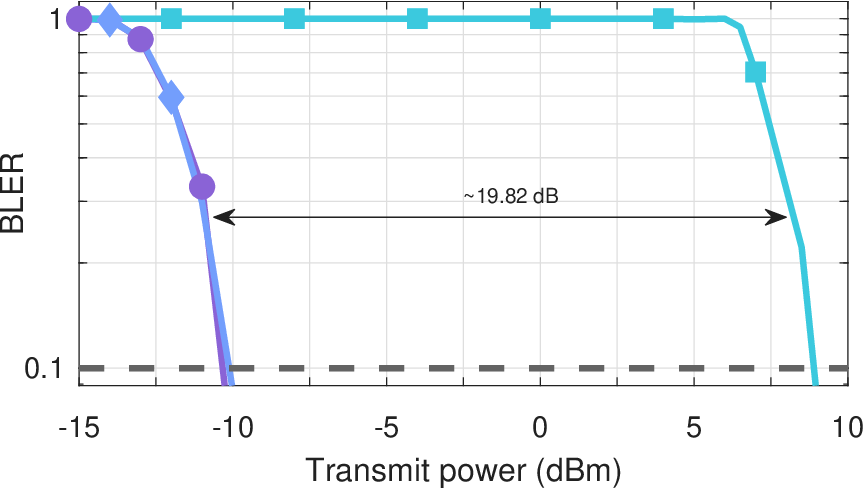}\label{fig:BLER_meanUE}
	\captionsetup{justification=raggedright,labelsep=period,singlelinecheck=false}
	\caption{\ Mean BLER for both UE devices as a function of the transmit power $P_\mathrm{Tx}$ for the scenario with $x_\mathrm{UE,1}=\SI{-495.33}{\meter}$ and $x_\mathrm{UE,2}=\SI{495.33}{\meter}$ ({\color[rgb]{0.5412,0.3922,0.8392}$\CIRCLE$}) and the scenario with $x_\mathrm{UE,1}=\SI{-742.99}{\meter}$ and $x_\mathrm{UE,2}=\SI{742.99}{\meter}$ ({\color[rgb]{0.4471,0.6196,0.9922}$\blacklozenge$}). For comparison, BLER curves for a single-UE scenario with both UE and satellite receiver at \mbox{$x=\SI{0}{\meter}$} and without \ac{SAC} \mbox{({\color[rgb]{0.2275,0.7922,0.8745}$\blacksquare$})} and a maximum tolerable BLER of 0.1 is highlighted \mbox{({\color[rgb]{0.3922,0.3922,0.3922}\rule[0.5ex]{0.4em}{1.5pt}\hspace{0.2em}\rule[0.5ex]{0.4em}{1.5pt}})} are also shown.}\label{fig:BLER}
	\vspace{-0.385cm}	
\end{figure}

The results in Fig.~\ref{fig:az_profiles} showed that resolving transmitting \ac{UE} devices is possible with $P_\mathrm{Tx}=\SI{-10}{dBm}$ for the considered system parameters, which is possible since the energy of all $M$ \ac{OFDM} symbol copies is compressed into the azimuth bin due to azimuth compression and \ac{DFT} processing, leading to a similar effect to \ac{SNR} processing gain in radar systems, though not coherent. To evaluate the effect of the transmit power on the overall communication performance, Fig.~\ref{fig:BLER} shows the achieved \ac{BLER} as a function of $P_\mathrm{Tx}$ for the same scenarios as for Fig.~\ref{fig:az_profiles}. The achieved results show that \mbox{$\mathrm{BLER}\leq0.1$} can be achieved with around \mbox{$P_\mathrm{Tx}\geq\SI{-10.33}{dBm}$} for both \ac{UE} devices in the first scenario due to ideal \ac{UE} separation in the angular domain. As for the second scenario, the aforementioned interference due to azimuth sidelobes results in a slightly higher transmit power requirement of \mbox{$P_\mathrm{Tx}\geq\SI{-10.12}{dBm}$} to achieve \mbox{$\mathrm{BLER}\leq0.1$}. Both scenarios present superior performance compared to the case without \ac{SAC} by around $\SI{20}{dB}$, which is due to the azimuth processing gain \mbox{$G_\mathrm{p,az}=\SI{19.68}{dB}$} traded off against a data rate redutction by a factor of \mbox{$M=93$}.





\section{Conclusion}\label{sec:conclusion}

This article introduced a novel coherent \ac{SAC} system concept for the \ac{DS2D} \ac{UL}. It involves a \ac{UE} transmitting copies of a signal that are processed at a satellite receiver to form a synthetic aperture. This allows the receiver to estimate the \ac{UE} direction, besides performing beamsteering and coherent combining. In addition to alleviating link budget requirements and enabling \ac{DoA} estimation of multiple \ac{UE} devices, the proposed \ac{SAC} system concept effectively manages interference via spatial multiplexing, making it well-suited for massive \ac{IoT} applications. Simulation results for a \ac{LEO} satellite with an orbit height of \mbox{$\SI{600}{\kilo\meter}$} and two \ac{UE} devices at \mbox{$\SI{3.5}{\giga\hertz}$} demonstrated that \mbox{$\mathrm{BLER}\leq0.1$} is achieved for resolved \ac{UE} devices in the angular domain for transmission powers of around $\SI{-10}{dBm}$ or higher.


\bibliographystyle{IEEEtran}
\bibliography{./References/references}

\end{document}